\documentstyle[aps,prb,floats,epsfig]{revtex}

\begin{document}
\draft

\title{Eu-Eu exchange interaction and Eu distribution in  Pb$_{1-x}$Eu$_x$Te
from magnetization steps}

\author{Ewout ter Haar, Valdir Bindilatti and Nei F. Oliveira Jr.}
\address{Instituto de F\'{\i}sica, Universidade de
S\~{a}o Paulo, C.\ P. 66318, CEP 05315-970 S\~{a}o Paulo, SP, Brazil}

\author{G.H. McCabe and Y. Shapira}
\address{
Department of Physics, Tufts University, Medford, Massachusetts 02155, USA}

\author{Z. Golacki}
\address{Institute of Physics, Polish Academy of Sciences, Pl. 02-668 Warsaw, 
Poland}

\author{S. Charar and M. Averous}
\address{Groupe d'Etude des semiconducteurs URA 357, Universit\'{e} 
Montpellier II, Place Eugene Bataillon, 34095 Montpellier Cedex 5, France}

\author{E.J. McNiff, Jr.}
\address{Francis Bitter National Magnet Laboratory, Massachusetts Institute of
Technology, Cambridge, Massachusetts 02139, USA}

\date{\today}
\maketitle
\begin{abstract}
The magnetization of Pb$_{1-x}$Eu$_{x}$Te samples with $x$ = 1.9, 2.6 and 
6.0\% was 
measured at 20~mK in fields up to 50~kOe, and at 0.6~K in fields up to
180~kOe. The 20~mK data show the magnetization steps (MSTs) arising 
from pairs and from triplets. The pair MSTs are used to obtain the 
dominant Eu-Eu antiferromagnetic exchange constant, $J/k_{B} = 
-0.264 \pm 0.018$~K.
The exchange constant for triplets is the same. Comparison of the 
magnetization curves with theoretical simulations indicates that the Eu 
ions are not randomly distributed over all the cation sites.  
The deviation from a random distribution is much smaller if $J$ is assumed 
to be the nearest-neighbor exchange constant $J_{1}$ rather than the 
next-nearest-neighbor exchange constant $J_{2}$.  On this basis, $J$ is  
tentatively identified as $J_{1}$.  However, the possibility that $J = J_{2}$ 
cannot be excluded completely. To obtain agreement with the data, it must be
assumed that the Eu ions tend to bunch together.
Comparision with microprobe data indicates that the length scale for these
concentration variations is smaller than a few $\mu$m. The theoretical 
simulations in the present work improve on those performed earlier by 
including clusters larger than three spins.  
\end{abstract}

\twocolumn

\section{INTRODUCTION}
\label{introduction}
An important group of dilute magnetic semiconductors (DMS) are lead salts in
which a fraction of the Pb ions have been replaced by Eu ions.\cite{1}   
Pb$_{1-x}$Eu$_x$Te is one member of this group. The other members are 
Pb$_{1-x}$Eu$_x$Se and Pb$_{1-x}$Eu$_x$S. All these materials have the 
rock-salt structure, with an fcc cation lattice. The Eu$^{2+}$ ion, with 
7 electrons in the half-filled 4$f$ shell, has zero orbital angular momentum
and total spin S=7/2. 
EPR data show that in Pb$_{1-x}$Eu$_x$Te $g = 1.98$ and that the spin 
hamiltonian for the Eu$^{2+}$ ion contains a very small crystal-field
anisotropy.\cite{2}
	
It has been known for several years that the Eu--Eu exchange interaction in 
these IV--VI DMS is two orders  of magnitude smaller than the Mn--Mn 
exchange interaction in the traditional II--VI DMS (e.g.,  Cd$_{1-x}$Mn$_x$Te).
In both types of DMS the exchange interaction is antiferromagnetic (AF), 
but the leading Eu--Eu exchange constant $J$ is of order $-0.1$~K, compared 
to $-10$~K for the Mn--Mn exchange constant. 

Much of the early information concerning these Eu--Eu exchange interactions 
came from measurements of the Curie-Weiss temperature $\theta$, and from 
analysis of high-field magnetization data at $T = 4.2$~K.\cite{3,4,5}
As discussed later, both of these methods yield only a rough estimate of 
the dominant AF exchange constant $J$. In the present paper we present an 
accurate determination of $J$ in Pb$_{1-x}$Eu$_x$Te using the 
magnetization-steps (MSTs)  method.\cite{6,7}  Because $|J|$ is of order 
0.1~K, the observation of the MSTs required the use of a dilution 
refrigerator operating well below 0.1 K. 
	
In addition to yielding $J$, the low-temperature magnetization data also gave 
information about the distribution of the Eu ions over the cation sites . 
Such information is not readily available by other means. 
The analysis indicates that the distribution is not perfectly random
(equal probability of occupation of all cation 
sites in the crystal). Instead, the Eu ions tend to bunch together. 
Another important issue is whether the dominant AF 
exchange constant $J$ corresponds to the nearest-neighbor (NN) or to the 
next-nearest-neighbor (NNN) exchange constant, $J_{1}$ or $J_2$ , 
respectively. This issue  is addressed on the basis of the results 
for the Eu spatial distribution.

\section{THEORY}
\label{theory}
Much of the relevant theory was discussed in a recent paper on the MSTs 
in Pb$_{1-x}$Eu$_x$Se.\cite{8}  We refer the reader to this earlier paper, 
and confine ourselves here to a summary of the main points. Some newer 
theoretical results are also mentioned.

As a first approximation we consider a simple model which captures the main 
features of 
the experimental data. In this ``single-$J$'' model for a DMS, only one AF 
exchange constant $J$ is 
included. Other exchange constants, and all anisotropies, are neglected. 
In this model each Eu ion on the fcc sublattice  belongs to a particular 
type of  ``cluster''.
The cluster types considered explicitly in Ref.~\onlinecite{8} are: 
singles (isolated magnetic ions which are not connected by any exchange bonds),
pairs, open triplets (OTs),  and closed triplets (CTs). More recently
the six types of quartets in the fcc lattice were considered also.\cite{9}

Assuming that $J$ is negative (antiferromagnetic), the magnetization
curve at low $T$ ($k_{B}T << |J|$) for clusters of a given type has two
main features: a quick saturation of the zero-field moment followed by a
series of MST's. \cite{8,9}
This behavior is illustrated by the exact calculation
of the magnetization curve for small clusters in this simple model. 
We will use normalized fields $h_{n} = g\mu_{B}H_{n}/|J|$ for the step
positions.
Pairs of Eu$^{2+}$ ions ($S=7/2$) have no zero-field moment. The seven MSTs 
from pairs occur at normalized fields $h_n = 2,4,...14$. Thus the pairs 
saturate at $g\mu_{B}H = 14|J|$. The next 
most important clusters are open triplets. They have a zero-field groundstate
with a net spin of 7/2 which saturates quickly. The MSTs occur at 
$h_n = 9,11...21$. Closed triplets have their MSTs at $h_n = 1,3...21$. 
Note that the last three MSTs due to triplets occur when the pairs are 
already saturated.  For five of the six types of 
quartets the series of MSTs end at 28$|J|$; only the ``string quartet'' 
ramp ends at a lower field $g\mu_{B}H = 24.2|J|$.\cite{9}

The exchange constant $J$ is usually obtained from the observed $H_n$ for
pairs since, except for singles, these are the most numerous. As was done 
in Ref.~\onlinecite{8},
we estimate an uncertainty in the determination of $J$ due to simplifications
introduced by the single-$J$ model. The cubic 
crystal-field anisotropy, the dipole-dipole interaction, and exchange 
constants other than $J$ basically shift and broaden the MSTs. Their
effects on the positions of the MSTs from pairs are included in the equation
\begin{equation}
\label{eq1}
g\mu_{B}H_{n} = 2n|J| + \Delta_{n}
\end{equation}
with $n = 1,2...7$. The shifts $\Delta_{n}$ depend
on $n$, the orientation of the sample (anisotropy) and the concentration 
(further neighbor exchange constants). We have calculated the 
magnetization and step postition of pairs taking into account the cubic 
crystal field and dipole-dipole interactions, using known parameters.\cite{2}
If $J$ is determined from a fit to Eq.~(\ref{eq1}), 
assuming a constant $\Delta_{n}$ (the procedure used below), 
the resulting error due to the crystal field
anisotropy is less than 5\%. The error due to dipole-dipole interactions is 
less than 1\%. 

Exchange interactions with further neighbors (not included in the
single-$J$ model will also shift and broaden 
the MSTs.\cite{12} These effects will be discussed later in connection
with the data analysis. 

The magnetization 
$M$ of the sample as a whole is obtained by adding the contributions of the 
various cluster types. The contribution of  each cluster type is the product 
of the magnetization per cluster   and the number of clusters of that 
type. To calculate the number of clusters of a given type one needs to know
how the magnetic ions are distributed over the cation sites. Normally, a 
random distribution is assumed. The populations of 
the various cluster types are then well known.\cite{9,10}  As discussed later,
deviations from a random distribution can be detected by analyzing the 
measured magnetization curve at low temperatures. The ability to detect such 
deviations is a major advantage. Any tendency of the magnetic ions to bunch 
together or to avoid each other is revealed. The determination of the AF exchange
constant $J$ is independent of the spatial distribution however.
	
In Ref.~\onlinecite{8} the theoretical simulations of the magnetization 
curves included clusters up to triplets.
In the present work the simulations were improved by adding the contributions 
of the six types of quartets.\cite{9} In addition a rough estimate of the 
contribution of clusters larger than quartets was also included, as will be
discussed shortly, so that all the spins were accounted for.  

Clusters larger than quartets (quintets, sextets, etc.) will be referred to 
collectively as ``others''. If the zero-field 
ground state of any such cluster has a net spin then this net spin will align rapidly 
with $H$ at low $T$. At higher fields a series of MSTs will occur. 
In practice, the MSTs from large clusters are very small in size, 
so that they are not resolved. The series of MSTs from a given cluster type 
then merges to form a ramp.  The ramp ends at a field which depends on the 
cluster type. 
	
The magnetization curve of the others is therefore a sum of many initial 
fast rises of $M$ followed by a superposition of many ramps ending at 
different fields. Here, we use an approximation in which the contribution
of the others is represented by a single initial fast rise of $M$ 
followed by one ramp. The initial fast rise is approximated by a 
Brillouin function for spin 7/2 with a saturation value corresponding to 
1/5 of the saturation value of the others. The 1/5 weight is motivated by 
earlier results for the initial rise of the magnetization.\cite{6}
The remaining magnetization rise
(4/5 of the saturation value of the others) is approximated by a 
single ramp which starts at $H = 0$ and ends at $g\mu_{B}H = 35|J|$. 
The latter value is expected to be slightly higher than the average 
saturation value for all quintets but lower than the average saturation 
value of sextets.

For our samples, the contribution of others to the 
magnetization was small, only 0.2\% of the saturation magnetization $M_0$  
for $x = 1.9\%$, and only 0.6\% of $M_0$ for $x = 2.6\%$. Even for the 
sample with $x = 6\%$ the others contributed only 9\% of $M_{0}$. 
Our approximation for the contribution of the others should therefore 
be adequate.

\section{EXPERIMENTAL TECHNIQUES}
\label{experiment}
The three Pb$_{1-x}$Eu$_x$Te samples were grown by the Bridgman method. 
The Eu concentration $x$ was determined from the saturation magnetization.
A moment of  $6.93\mu_{B}$  
per Eu ion (based on $S = 7/2$ and $g = 1.98$) was assumed. A small correction 
for the lattice susceptibility, $\chi_{d} = -3\times 10^{-7}$~emu/g, 
was applied.\cite{3}  The results for the three samples gave 
$x$ = 1.9, 2.6 and 6.0\%. These values are supported by the Curie constants, 
obtained from susceptibility data, which gave $x$ = 1.9, 2.4, and 5.9\%, 
respectively. We shall adopt the first set of values, which we regard as 
more accurate.  
	
The Eu concentration $x$ was also obtained from microprobe measurements. 
Approximately 35 spots on a single surface of each of the samples were probed. 
Each spot had a diameter of 5~$\mu$m, and probed the Eu concentration to a 
depth of about 2~$\mu$m. The average values for $x$ and the standard 
deviations were $1.7 \pm 0.3\%,  2.5 \pm 1.1\%$, and $5.6 \pm  0.2\%$. 
The large standard deviation for the second sample  is due to a small region 
of higher concentration near one corner.  The other two samples, 
particularly the one with the highest concentration, were quite homogeneous 
on length scales larger than several $\mu$m.
	
Magnetization measurements at 20~mK were made using a force magnetometer 
which operated in the mixing chamber of a plastic dilution refrigerator. 
This equipment was described earlier.\cite{7,8,13,14}  The main magnetic 
field $H$, up to 50~kOe, was generated by a superconducting magnet. 
The force was produced by a superimposed dc field gradient, 
$dh/dz = 0.8$~kOe/cm,
which was generated by an independent set of superconducting coils. 
With this gradient the variation of the field over the volume of the sample 
was less than 0.2~kOe.  None of the samples were oriented, so that the 
direction of the field relative to the crystallographic axes was not known. 
As discussed later, the anisotropy in Pb$_{1-x}$Eu$_x$Te  is small, so 
that field orientation is not critical.
	
The magnetization 
was also measured in fields up to 180~kOe using a vibrating sample 
magnetometer (VSM) which was adapted for use in  a Bitter magnet. 
The samples were immersed in liquid $^3$He at 0.6~K. Other magnetization data
were taken with a SQUID magnetometer system \cite{14a} at
temperatures above 2~K and in fields up to 50~kOe. Among the data taken 
with this system were the low-field susceptibility 
data used to obtain the Curie constant and  the Curie-Weiss temperature
$\theta$.

\section{RESULTS AND DISCUSSION}
\label{discussion}
\subsection{Magnetization curves}
\label{magcurves}

\begin{figure}[bt]
\centerline{\epsfig{file=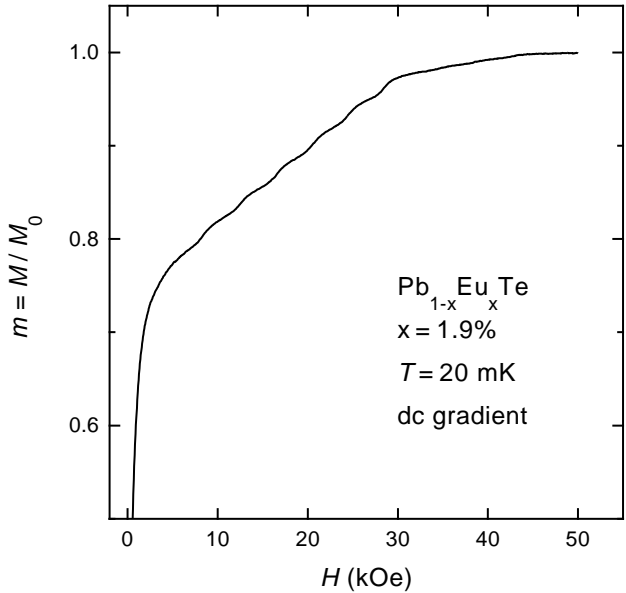,width=\linewidth}}
\caption{\label{fig1} Top half of the magnetization curve at 20~mK for the 
sample with $x$ = 1.9\%. The magnetization $M$ has been normalized to its 
saturation value $M_0$.}
\end{figure}

\begin{figure}[bt]
\centerline{\epsfig{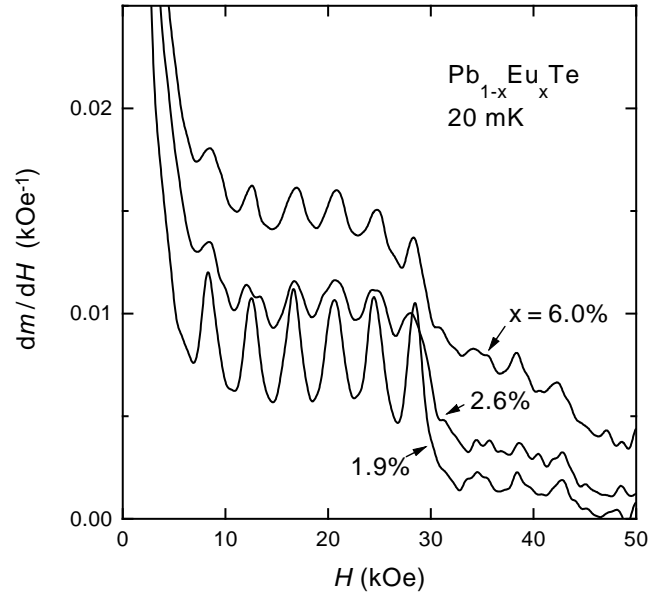}}
\caption{\label{fig2} Field  derivative of the normalized magnetization 
$m = M/M_0$ for the three samples at 20~mK. These results were obtained 
by a numerical differentiation of the magnetization curves.}
\end{figure}

Figure~\ref{fig1} shows the top half of the magnetization curve at 20~mK for 
$x = 1.9\%$. The fast rise at the lowest fields is due to the alignment of 
singles, with minor contributions from other clusters with zero-field moments.
This initial 
fast rise of $M$ is followed by a large ramp on which six MSTs are clearly 
visible. The six MSTs are due to pairs. Only six, instead of 
seven, MSTs are visible because the first MST is masked by the large initial 
fast rise of $M$ on which it is superimposed. The ramp due to pairs ends 
near 28~kOe. It is followed by a much less steep ramp due to open triplets 
(OTs) which ends near 43~kOe. (The OT ramp is predicted to start well before 
the pair ramp ends, but because the pair ramp is so much steeper the OT ramp 
does not stand out in this field region.)  For this sample, with the lowest 
concentration, the magnetization becomes practically saturated once the OT 
ramp ends.

The lowest curve in Fig.~\ref{fig2} shows the differential susceptibility 
(field derivative of the magnetization) obtained numerically from the data in 
Fig.~\ref{fig1}. The six MSTs due to pairs appear as large peaks. Three much 
smaller peaks are seen between 34 and 43~kOe. These are the last MSTs from the 
OTs. To our knowledge this is the first clear observation of MSTs from 
triplets in a DMS.  The other two curves in Fig.~\ref{fig2} show similar 
results for pair and triplet MSTs in the other two samples.

\begin{figure}[bt]
\centerline{\epsfig{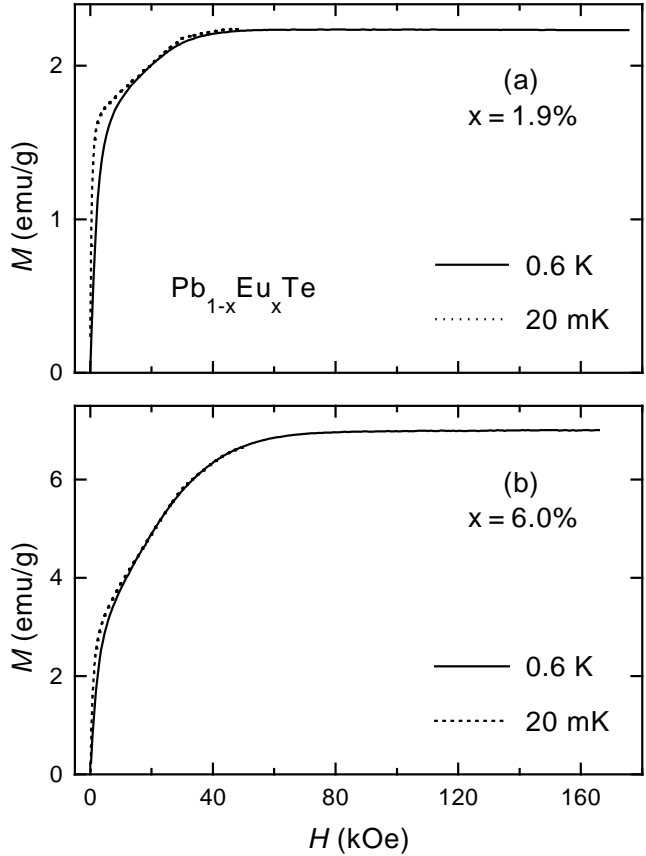}}
\caption{\label{fig3}Magnetization curves for $x$ = 1.9 and 6.0\% at 0.6~K. 
These results have been corrected for the lattice diamagnetism. Also shown 
are the  magnetization curves at 20~mK, with the vertical scale adjusted 
so that they match the 0.6~K data at 50~kOe.}
\end{figure}
	
Figure~\ref{fig3} shows magnetization curves for two of the samples at 0.6~K. 
These curves extend to fields of about 180~kOe. Also shown, for comparison, 
are the corresponding magnetization curves at 20~mK, adjusted so that they 
agree with the 0.6~K data at 50~kOe.  The main differences between the 0.6~K 
and 20~mK data are that at 0.6~K the initial rise of $M$ is more gradual, 
the pair ramp is more rounded, and the  pair MSTs are not resolved. 
These differences are the expected effect of temperature.
The data at 0.6~K show that for $x = 1.9\%$ complete saturation is achieved 
near 60~kOe.  For $x = 6.0\%$ however, the approach to saturation is
more gradual because a larger percentage of the spins are in large clusters 
which saturate more slowly.


\subsection{Dominant AF exchange constant $J$}
\label{dominant}

\begin{figure}[bt]
\centerline{\epsfig{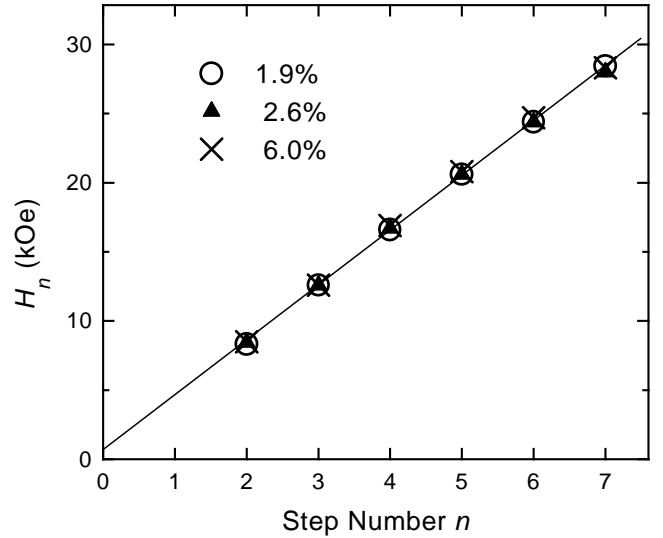}}
\caption{\label{fig4} The fields $H_{n}$ at the MSTs  from pairs as a 
function of the step number $n$. The straight line represents Eq.~(\ref{eq1})
with the average values for $J$ and $\Delta_n$.}
\end{figure}

In all three samples the pair and triplets involving the exchange constant
$J$ gave rise to distinct ramps on which well-resolved MSTs were observed.
No additional ramps or MSTs from any other exchange constant were found. 
The single-$J$ model therefore seems to be appropriate for describing
the magnetization curves. Additional evidence that the corrections for the 
single-$J$ model are small will be discussed later.

The fields $H_{n}$ at the pair MSTs were obtained from the peaks in 
Fig.~\ref{fig2}.  Figure~\ref{fig4} shows a plot of these fields for the 
three samples. Because the first MST was not observed, the plot is only for 
$n = 2$--7.  It is evident that $H_{n}$ is approximately linear in $n$,
so that Eq.~(\ref{eq1}) with a constant $\Delta_{n}$ is a good approximation
(see Ref.~\onlinecite{15}).

The values for $J$ obtained  from least 
squares fits to Eq.~(\ref{eq1}), holding $\Delta_n$  constant, were the same 
for all three samples within 2\%. The average value was $J/k_{B} = -0.264$~K.
The values for $\Delta_{n}$  obtained from fits of the three data sets 
to Eq.~(\ref{eq1}) were all positive. The average was $\Delta_{n} = 0.69$~kOe,
which is only a fraction of the spacing $\Delta H = 4$~kOe between 
successive MSTs.

As far as further neigbor exchange interactions are concerned, 
we can say the following.
The data show that any other AF exchange constant is at least a factor 
of 3 smaller than the observed $J$. If this were not the case, another
noticeable ramp ending between 10~kOe and 30~kOe should have been present, 
leading to a change of slope in this field region. Such a change of slope was 
not observed. Furthermore, the observed MSTs show only a small amount of 
non-thermal broadening and $\Delta_{n}$ depends only weakly on $x$. 
These observations show that the magnetization above about 10~kOe is very 
little affected by other (smaller) exchange constants originating
from further neigbors.

The following arguments indicate that the quoted value $J/k_{B} = -0.264$~K
corresponds to the largest AF exchange constant. The 0.6~K data in 
Fig.~\ref{fig3} show no other MSTs or ramps in fields up to 175~kOe. This 
means that if there were any larger AF exchange constant, its magnitude 
should have exceeded 12~K. On the other hand the Curie-Weiss temperatures 
discussed below rule out such a large AF exchange constant. Thus, the observed 
MSTs gave the largest $J$. 

Our final result is then that the dominant exchange constant has the
value $J/k_{B} = -0.264 \pm 0.018$~K.
The 7\% uncertainty is motivated by the calculations of the corrections
mentioned in Sec.~\ref{theory}, and the observation that further neighbor
interactions affect the MSTs very little.
This $J$ for Pb$_{1-x}$Eu$_x$Te is only slightly larger than $-0.24$~K for 
Pb$_{1-x}$Eu$_x$Se.\cite{8}  There was no measurable dependence of $J$ on 
$x$ in the present Pb$_{1-x}$Eu$_x$Te samples (see Fig.~\ref{fig4}), 
despite the fact that the band gap changes by nearly a factor of 2 as $x$ 
changes from 1.9 to 6.0\%.\cite{1}
The fields at the three observed MSTs from OTs, above 32~kOe, are close to 
those expected from the value of $J$ derived from the pair steps.  
This means that the exchange constants for pairs and triplets are the same 
(within a few percents), as expected.

Low-field susceptibility data were taken at temperatures 
down to 2~K.  The most accurate result for $\theta$ was for the sample with 
$x = 6.0$\%. With the usual assumptions \cite{15a} the value $\theta = -1.9$~K 
gave $J/k_{B} = -0.25$~K (assuming $J=J_1$). 
The results for the other two samples, with lower 
$x$, were $J/k_{B} = -0.30$ ($x = 2.6\%$) and $-0.22$~K ($x = 1.9\%$), 
but these were judged to be less accurate. 
The values found  by Gorska {\sl et al.} \cite{3}  were $J/k_{B} = -0.38$~K 
for $x = 3\%$, and  $-0.27$~K for $x = 6\%$. The spread in the values of $J$ 
obtained from $\theta$ shows that that the experimental 
uncertainty is not negligible, especially for low concentrations.  
Determining $J$ from the Curie-Weiss $\theta$ has several other drawbacks. 
First, it assumes that the distribution of the Eu ions is random, which, as 
discussed later, may not be exactly true for Pb$_{1-x}$Eu$_{x}$Te. Second, 
$\theta$ depends on all exchange constants. Third, even when one $J$ is much 
larger than all the others, its estimate depends on
whether it is identified as $J_1$ (between NNs, as was done above) 
or $J_2$ (between NNNs). 
The identification of $J$ as $J_2$ would lead to an estimate of $J$ 
which is larger by a factor of 2 compared to $J=J_1$.  
The MSTs method is a direct determination of $J$, which is independent of the 
identification of $J$, and is also independent of the spatial distribution 
of the Eu ions.
	
Gorska {\sl et al.} \cite{3} also determined $J$ by analyzing the 
magnetization curve at 4.2~K.  At this relatively high temperature,  
$k_{B}T = 16|J|$, the MSTs are not resolved, and the ramps due to pairs and 
OTs do not stand out clearly. The analysis which was performed fitted the 
magnetization curve to a sum of two contributions: one from singles and the 
other from ``PAIRS''. The value for $J$ was deduced from the PAIR contribution.
The results were $J/k_{B} = -0.43$~K for $x = 3\%$, and  
$-0.50$~K for $x = 6\%$.
The reason that these values are much higher than our value of $-0.264$~K,
is that the assumed ``PAIRS'' actually included not only true pairs but also 
triplets, quartets, and larger clusters. Since the saturation field of a 
cluster increases with cluster size, the net effect was that the assumed 
PAIRS saturated at higher fields than true pairs. This caused the deduced $J$ 
to be higher than the true $J$.  It is probably significant that the fitted 
$J$ was larger for $x = 6\%$ than for $x = 3\%$, because the percentage of 
large clusters increases with $x$.

\subsection{Eu distribution and the identity of $J$}
\label{distribution}

\begin{figure}[bt]
\centerline{\epsfig{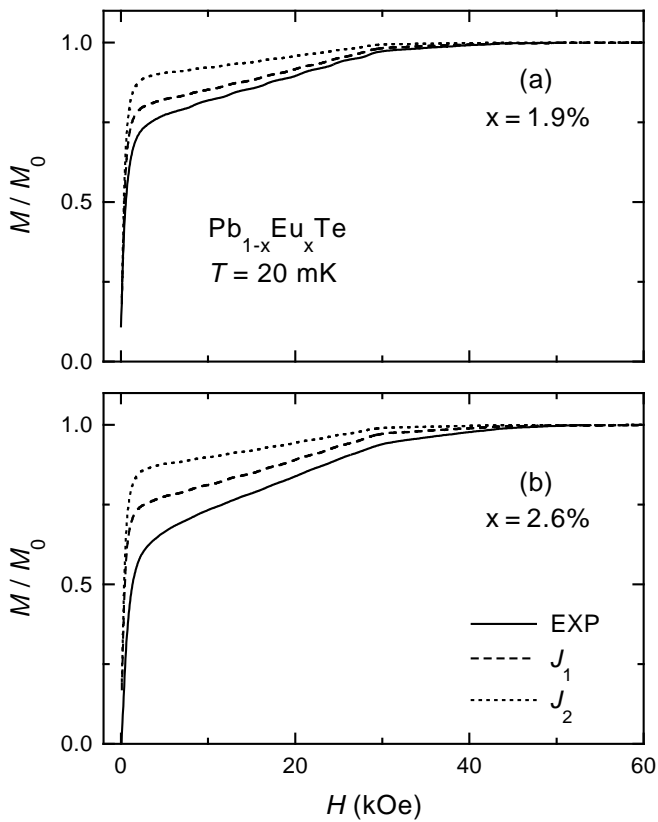}}
\caption{\label{fig5}Comparison between the measured magnetization curve at 
20~mK with computer simulations based on the $J_1$ -  and $J_2$ models. 
a) $x$ = 1.9\%.  b) $x$ = 2.6\%.}
\end{figure}

\begin{figure}[bt]
\centerline{\epsfig{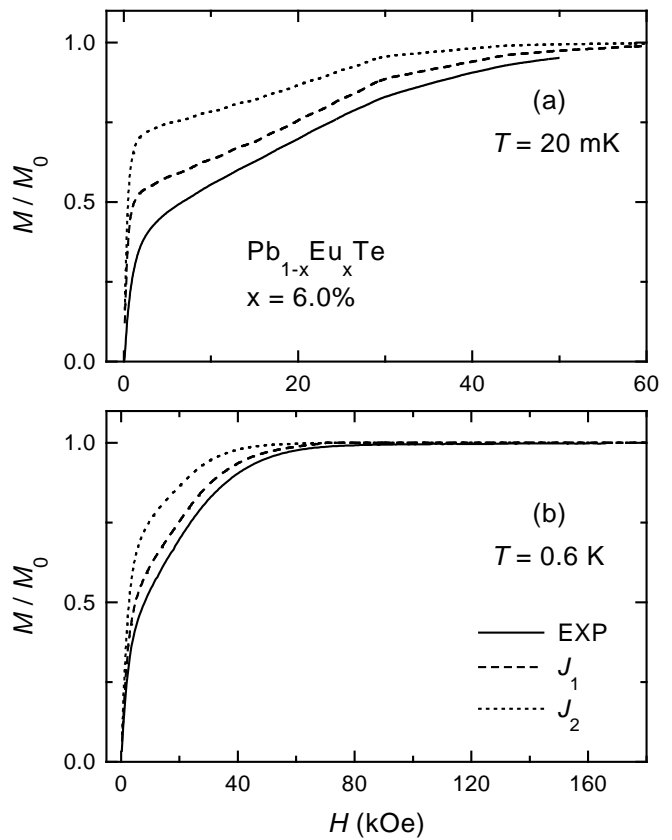}}
\caption{\label{fig6} Comparison between the measured magnetization curve 
for $x$ = 6.0\% and computer simulations based on the $J_1$ -  and $J_2$ 
models. a) $T = 20$~mK,  b)  $T = 0.6$~K.}
\end{figure}

The identity of the dominant AF exchange constant $J$, whether it is $J_1$ 
for NNs or $J_2$ for NNNs, is a significant issue. Normally, exchange 
constants decrease rapidly with distance, so that the dominant $J$ is 
between NNs. This expected normal behavior has been found in all 
Mn-based II--VI DMS.\cite{6}  On the other hand, in pure EuTe the 
largest exchange 
constant is $J_2$, which is antiferromagnetic with a value of about $-0.2$~K.  
The exchange constant $J_1$ in EuTe is ferromagnetic and smaller in 
magnitude.\cite{16,17} Thus, after $J$ was measured in the present work, 
it was still not obvious whether it was $J_1$ or $J_2$. 
	
To address this problem the measured magnetization curves were compared with 
computer simulations based on the two competing hypotheses: $J = J_1$
($J_1$ model), or $J = J_2$ ($J_2$ model). Both models are single-$J$ models. 
The simulations were similar to those in Ref.~\onlinecite{8} except that 
they also included the quartets and the others. We also chose to normalize the 
magnetization curves to the saturation value $M_{0}$ rather than to the data 
point at the highest field (see Ref.\onlinecite{18}). The key assumption in 
the simulations was that the Eu ions were distributed randomly, i.e., 
the probability of occupation of each cation site in the crystal was $x$. 
The simulations with the $J_1$ - and $J_2$ models lead to different 
magnetization curves, essentially because there are 12 NNs but only 6 NNNs in 
the fcc cation lattice. This difference means that in the $J_2$  model the 
number of singles is higher and the populations of pairs and larger clusters 
are lower. 
	
Figure~\ref{fig5} compares the magnetization curves at 20~mK, for $x = 1.9$
and 2.6\%, with simulations based on the two models. To account for the 
somewhat larger observed width of the MSTs than that expected from thermal 
broadening alone, the simulations of data at 20~mK used an effective 
temperature of 100~mK. This change has no effect on the discussion below. 
\cite{8} The simulation of data at 0.6~K were made with the actual 
temperature.
From the figure it is clear that the $J_1$  model simulation is closer to the 
data, but even 
this model fails to reproduce the measured curves exactly. The deviations are 
more pronounced for $x = 2.6$\%. The observed initial rise of $M$ is smaller 
than predicted,  indicating that even the $J_1$  model overestimates the 
number of singles. 
Furthermore, the observed slope between 30 and 40~kOe is higher than 
predicted by both models. This means that there are more triplets 
and/or larger clusters than calculated using a  random distribution. 
The same features are also seen in Fig.~\ref{fig6}, which is for $x = 6.0\%$ 
at 20~mK and at 0.6~K. The $J_1$ model is closer to the data, but even 
compared to this model the actual number of singles is lower and the number 
of triplets and/or larger clusters is higher.

These results indicate that the {\em local} Eu concentration in the vicinity 
of a typical Eu ion is higher than the average concentration for the sample 
as a whole. Thus, the Eu ions tend to 
bunch together, leading to an inhomogeneous  Eu distribution. 
The size of the regions with higher Eu concentration cannot be inferred from  
analysis of the magnetization curves alone. The microprobe measurements 
(Sec.~\ref{experiment}) examined the concentration variations on length scales
greater than a few $\mu$m. The  attempt to account for the magnetization 
curves by using the microprobe results as the local concentration profile was 
unsuccessful. This is particularly evident for the sample with $x = 6.0\%$,  
in which the microprobe concentration showed very little variation with 
position. Combining the analysis of the magnetization data with the 
microprobe results we conclude that the length scale of the concentration 
inhomogeneities was smaller than a few $\mu$m. 
	
A quantitative measure of the degree to which the Eu ions bunch together can 
be obtained by introducing the concept of a local concentration $x_{L}$. 
The simple picture which is behind this concept presumes that there is a 
typical local concentration in the vicinity of a typical Eu ion. It can 
differ from the average concentration $x$ for the sample as a whole. 
The populations of the various clusters are calculated assuming that the 
probability of occupation of each cation sites in the vicinity of a Eu ion 
is $x_L$   instead of $x$. The value of $x_L$  is found by varying the Eu 
concentration in the simulation until a match with the experimental 
magnetization curve is achieved. Using this procedure the following results 
were obtained for the samples with $x$ = 1.9, 2.6 and 6.0\%. With the $J_1$
model, $x_L$ = 2.5, 3.8, and 7.7\%, respectively. With the $J_2$ model, 
$x_L$ = 4.7, 7, and 14\%, respectively. Obviously, a much smaller difference 
between $x_L$ and $x$ is required to achieve agreement with the $J_1$ model. 

In the discussion above, we have assumed that the deviations of the data
from a single-$J$ model were due to the assumption of a random distribution.
An alternative would be to question the validity of the single-$J$ model
itself. It is known that long range exchange interactions between
further neighbors can influence the 
magnetization process in diluted magnetic 
semiconductors.\cite{12,denissen86,twardowski87,vu92} 
Such long-range interactions may be important in small-gap DMS, 
such as the present system. The key question here is whether long-range 
interactions could have affected the magnetization in fields above 10 kOe 
significantly, since the conclusion of a non-random distribution was based 
on the data at these high fields, especially above 30~kOe. 

The effects of further neighbors (distant neighbors) on the magnetization curve 
have been treated using several approximate methods. One approach starts from the 
clusters in the single-J model, but then subjects these clusters to effective fields 
arising from further neighbors.\cite{12} Alternative approaches use more general 
spin clusters which include further neighbors within the clusters, so that the intracluster 
interactions already include the further neighbors.\cite{denissen86,twardowski87,vu92} 
Since our analysis started from the single-J model, the first approach 
(effective fields acting on the single-J clusters) is more convenient here. 
The effective fields from further neighbors slow down the alignment of spins which are 
singles, and they shift and broaden the ramps and MSTs arising from larger clusters 
(e.g. pairs and triplets). These effects increase with the Eu concentration $x$. 
We now present several arguments which indicate that in the present work the 
further-neighbor effective fields were far too weak to significantly affect the 
magnetization curve well above 10 kOe.

At 20~mK the magnetization rises very quicly at low $H$ (Figs.~\ref{fig1} and \ref{fig5}). 
The singles seem to become practically saturated at about 2~kOe. Considering that the 
cubic crystal field anisotropy also slows down the alignment of the singles,\cite{8} 
we conclude that typical further-neighbor effective fields acting on singles are less than 
about 1~kOe. Such weak effective fields are unlikely to have a significant effect on the 
magnetization well above 10~kOe. 

The fields $H_n$ at the MSTs from pairs are shifted by the further-neighbors 
effective fields.\cite{12}  These shifts, included in the $\Delta_n$ of Eq.~(\ref{eq1}), 
are predicted to increase with $x$. However, the results in Fig.~\ref{fig4} indicate 
that the $\Delta_n$ are all less than about 1~kOe. Also, despite the factor-of-three 
change in $x$, the fields $H_n$ do not vary by more than a fraction of 1~kOe. 
This means that typical further-neighbor effective fields acting on pairs were no more than 
a fraction of a kOe, even for the sample with $x = 6$\%. Typical effective fields 
acting on triplets and quartets should not be higher by more than a factor of 2 or so. 
It is very unlikely that such weak effective fields will have a significant effect on 
the magnetization curve well above 10 kOe.

The {\em spread} in the magnitude of the effective field acting on different pairs gives rise 
to a broadening of the MSTs from pairs.\cite{12} Such a broadening should increase 
with $x$. The fact that well resolved MSTs were observed even for $x = 6$\% means 
that the spread in the effective fields was considerably smaller than the 4~kOe separation 
between the MSTs. Thus not only was the average effective field small, but the spread 
was also small. Such a distribution of effective fields should not have a significant 
effect on the magnetization curve well above 10~kOe.

The final argument in support of a non-random distribution is based on an analysis of 
the change in the slope of the magnetization curve near 43~kOe. This change occurs 
when the triplet ramp ends. (The change in slope is visible in 
Figs.~\ref{fig1}~and~\ref{fig5}, but is clearer in expanded plots of the magnetization 
curves at 20~mK.) The magnitude of this change in slope is related to the number of 
triplets. Analysis of the data in all three samples indicates a significantly larger 
number of triplets than predicted assuming a random distribution. 
Further-neighbor interactions cannot account for the significanly larger change in 
slope which was observed.

Claims of a non-random distribution in II-VI DMS have been made more than a 
decade ago (see e.g. Ref.~\onlinecite{galazka80}). However, MST studies showed 
that the distribution in such DMS was in fact random.\cite{shapira84,6}
Here, in our Pb$_{1-x}$Eu$_x$Te samples, we 
find a small non-randomicity which we believe to be genuine. 
In Ref.~\onlinecite{8} (Pb$_{1-x}$Eu$_x$Se)
deviations of the data with the $J_1$-model were also found, but the 
simulations included only clusters up to 
triplets. A re-analysis of those data with our improved model, which includes
quartets and an approximation for larger clusters, still does not lead
to perfect agreement. As in the present work, 
good agreement with the data is obtained if small deviations from a random 
distribution are assumed (and that $J = J_1$). 
The non-randomicity in the Pb$_{1-x}$Eu$_x$Se samples
is smaller than that in the Pb$_{1-x}$Eu$_x$Te samples.
	
Returning to the issue of the identity of $J$ , there are two possibilities. 
Either $J$ is $J_1$ in  which case the Eu ions have only a fairly modest 
tendency to bunch together, or $J$ is $J_2$  with a very large increase of the 
local Eu concentration.  Because the usual distribution of magnetic ions in a 
DMS is random,\cite{6} we believe that the first possibility is more likely. 
On this basis  we tentatively identify $J = -0.264 \pm 0.018$~K as $J_1$. 
However, since the actual distribution is unknown, the possibility 
that it is $J_2$   cannot be excluded entirely. If one accepts that $J$ is 
$J_1$ then the exchange constants in Pb$_{1-x}$Eu$_x$Te when $x$ is 
low are very different from those in EuTe.\cite{16,17}  Such a difference 
may be the result of a different band structure and a different position of 
the Eu levels.

\section{Acknowledgments}
We  are grateful to C. Merlet of the University of Montpellier II for 
assistance in the microprobe measurements, and to M.T. Liu for help in some of 
the magnetization measurements.  The work in Brazil was supported  by CNPq, 
FAPESP, and FINEP.  The work in the U.S. was partially supported by NSF 
Grants Nos. DMR-9219727 and INT-9216424. The work in France was supported by 
CNRS. The work in Poland was supported by the Polish Committee for Scientific 
Research. The Francis Bitter National Magnet Laboratory was supported by NSF.


\begin{references}
\bibitem{15} Detailed examination of the observed spacings $\Delta H_{n}$ 
between successive MSTs shows that, for each of the samples, these spacings 
vary relative to the mean value by up to 10\%. Some of this variation is due 
to experimental uncertainties but the main cause is probably the cubic 
crystal field  anisotropy. 
In that case, $\Delta_{n}$ in Eq.~(\ref{eq1}) is not exactly independent of $n$. 
Nevertheless, making a fit with constant $\Delta _{n}$ should give quite an 
accurate value for $J$.
\bibitem{18} The only significant difference between the two  normalization 
procedures was for $x = 6.0\%$ at 20~mK, because $M$ was not saturated at 
the maximum field of 50~kOe. The normalization in this case was based 
on the good match between the data at 20~mK and 0.6~K for fields above 15~kOe, 
as shown in Fig.~3(b). It was assumed that this good match persisted above
50~kOe.
\end{references}
\end{document}